\documentstyle[prl,aps,preprint,epsf,rotate]{revtex}
\newcommand{\eq}{\begin{equation}}
\newcommand{\ee}{\end{equation}}
\newcommand{\eqa}{\begin{eqnarray}}
\newcommand{\eea}{\end{eqnarray}}

\def\sxy{\sigma_{xy}}
\def\sxx{\sigma_{xx}}
\def\rxx{\rho_{xx}}

\def\s{\sigma}

\def\dg{\delta G}

\def\rxx{\rho_{xx}}

\begin{document}
\draft
\title{Critical Conductance and Its Fluctuations at Integer Hall Plateau
  Transitions}
\author{Ziqiang Wang$^{a,b}$, Bo\v zidar Jovanovi\'c$^b$, and Dung-Hai Lee$^c$}
\address{$^a$Department of Physics, Boston College, Chestnut Hill,
              MA 02167}
\address{$^b$Department of Physics, Boston University, Boston, MA 02215}
\address{$^c$Department of Physics, University of California at
  Berkeley, Berkeley, CA 94720}

\date{\today}
\maketitle

\begin{abstract}
Under periodic boundary condition in the transverse direction, 
we calculate the averaged zero-temperature two-terminal conductance 
($<G>$) and its statistical fluctuations
($<(\dg)^{2n}>$ for $n\le 4$) at the critical point of integer quantum 
Hall plateau transitions. We find {\it universal} values for 
$<G>=(0.58\pm0.03){e^2\over h}$, and 
$<(\dg)^{2n}>=({e^2\over h})^{2n}A_{2n}$, where 
$A_{2,4,6,8}=0.081\pm0.005$; $0.013\pm0.003$; $0.0026\pm0.005$;
and $(8\pm2)\times10^{-4}$ respectively. We also determine the 
leading finite size scaling corrections to these observables.  
Comparisons with experiments will be made.
\end{abstract}

For macroscopic, disordered, two dimensional electronic systems at 
low temperatures, the metallic behavior is only 
observed near quantum phase transitions. 
Two examples are the superconductor to insulator transitions 
in amorphous thin films, and the transitions between quantum Hall
plateaus \cite{reviewbook}. In two spatial dimensions,
the conductivity tensor $\s_{\mu\nu}$ measured in units of $e^2/h$ 
is dimensionless. Under generic conditions,  $<\s_{\mu\nu}>$ 
(the impurity averaged conductivities) are expected to be universal 
at the quantum critical points \cite{fisher,klz}.

The universal conductance fluctuations of disordered mesoscopic metals 
has attracted tremendous experimental and theoretical interests 
in recent years \cite{leestone}. The same question can be asked at 
the quantum critical points mentioned above. Thus 
``what is the statistical (over the impurity ensemble) properties 
of $\s_{\mu\nu}$ at quantum critical points of two dimensional systems''
is the concern of the present paper. In particular, we concentrate on 
the transitions between integer quantum Hall plateaus. 

In an integer plateau transition, the electron conductivity tensor 
$(\sxx,\sxy)$ changes from $(0, n){e^2\over h}$ to $(0,n\pm1)
{e^2\over h}$. Concerning these transitions the following consensus are 
reached \cite{reviewbook,wei,koch,wei2,engel}.
(a) When extrapolated to zero temperature and infinite sample size, 
the transitions are genuine continuous phase transitions with 
a {\it single} divergent length scale $\xi$ - the quasiparticle 
localization length. (b) As the Fermi energy ($E_F$) moves toward a 
critical value $E_c$, $\xi\sim\vert E_F-E_c\vert^{-\nu}$, with 
$\nu\simeq2.3$. (c) The characteristic 
energy scale $\sim 1/\xi$, thus (the dynamical exponent) $z =1$.

Based on a Chern-Simons formalism, Kivelson, Lee, and Zhang (KLZ) 
{\it  asserted} that $(\sxx,\sxy)$ at the $(0,n){{e^2}\over h}\to
(0,n+1){{e^2}\over h}$ transition
is $(\sxx^c,\sxy^c)=({1\over 2},n+{1\over 2}){{e^2}\over h}$
\cite{klz}. Huo, Hetzel, and Bhatt have numerically computed $\sxx^c$ at the 
$(0,0)\rightarrow (0,1){{e^2}\over h}$ transition using the Kubo
formula \cite{huo}.
Their calculation assumes that the electronic states lie entirely in the
lowest Landau level. They have considered several different
forms of the impurity potential and concluded that $\sxx^c=(0.55\pm0.05)
{e^2\over h}$ which is consistent with the result of KLZ.

Recently, the experimental four-terminal resistivity $\rxx$ has been studied
systematically at the quantum Hall liquid to insulator transitions
by Shahar {\it et. al.}\cite{shahar}.
They concluded that the critical conductivity tensor is {\it universal} and
the value of $\rxx^c\approx h/e^2$ agrees with the prediction KLZ.
In addition, significant sample to sample fluctuations in the critical
resistance have been observed \cite{shahar}.
Here we note that in order to measure the true $\rxx^c$ (or $\sxx^c$), 
the temperature has to be low enough so that corrections to scaling have
faded away, yet it has to be high enough so as to avoid the effects of 
mesoscopic fluctuations \cite{footnote}. 
More recently, the statistical fluctuation of the
two-terminal conductance in the transition regime 
has been studied experimentally in mesoscopic samples
by Cobden and Kogan who demonstrated
the presence of large mesoscopic fluctuations in the conductance 
\cite{cobden}.

In this paper, we calculate the ensemble averaged {\it two-terminal} 
conductance $<G>$ and, for the first time, its fluctuations
$<(\delta G)^{2n}>$  at the critical point of the integer quantum Hall 
plateau transitions.
In addition, we study the finite-size and aspect 
ratio dependence of these quantities.
The model we use is the Chalker-Coddington network model \cite{cc}
and its extension \cite{lwk}, and 
with periodic boundary conditions applied in the transverse direction. 
These models have been shown to exhibit the
correct critical properties of the integer plateau transitions 
\cite{cc,lwk,lw1}. Our findings are summarized as follows. 
The conductance of a $W\times L$ sample (where $W$ is the circumference 
and $L$ is the length of the cylinder) exhibits the following scaling behavior,
\eq
<G>={{e^2}\over h}{\cal F}(L^{y_{\rm rel}}\Delta g_{\rm rel},
L^{-y_{\rm irr}}\Delta g_{\rm irr}, W/L).
\label{sigma}
\ee
Here we have used the fact that $G/({e^2\over h})$ is dimensionless. In the
above, $<...>$ denotes the disorder ensemble averaging, $\Delta g_{\rm rel}$
and $\Delta g_{\rm irr}$ are the coupling constants conjugate to the 
relevant and the least-irrelevant operators respectively. 
Thus $y_{rel}\equiv 1/\nu$ and
$-y_{\rm irr}$ are the renormalization group dimensions of 
$\Delta g_{\rm rel}$ 
and $\Delta g_{\rm irr}$ respectively. At the critical point, 
$\Delta g_{\rm rel}=0$. We have performed finite size
scaling analysis of $<G>_c$ on $L\times L$ samples to extract 
${\cal F}(0,0,1)$ and $y_{irr}$ to be $(0.58\pm0.03)$ and $0.55\pm0.15$ 
respectively.  Thus the critical conductance $<G>_c=(0.58\pm0.03)
{{e^2}\over h}$. We have also calculated the central moments
$<(\dg)^{2n}>$ for $n\le 4$ 
and shown that they exhibit the following scaling behavior,
\eq
<\dg^{2n}>=({{e^2}\over h})^{2n}{\cal F}_{2n}(L^{y_{\rm rel}}
\Delta g_{\rm rel}, L^{-y_{\rm irr}}\Delta g_{\rm irr}, W/L).
\label{sigma2}
\ee
We have determined the critical moments $<\dg^{2n}>_c=(e^2/h)^{2n}{\cal
F}_{2n}(0,0,1)$. For $n=1,2,3,4$, the values of ${\cal F}_{2n}$ 
are found
to be $0.081\pm0.005$, $0.013\pm0.003$, $0.0026\pm0.005$, 
and $(8\pm2)\times10^{-4}$ respectively.
We verified the universality of these results and assert that all higher
moments, and hence the {\it entire} distribution function $P(G)$, 
is universal at the transition.
Since in the rest of this paper we {\it ignore} the electron-electron 
interaction, our conclusions are at most relevant to the {\it integer} 
plateau transitions. We also point out that 
it remains {\it unclear} at present about the relation 
between the two-terminal conductance $G$ and the $\s_{\mu\nu}$ derived
from the Kubo formula in a closed system without contacts.

Let us consider {\it non-interacting} electrons in a strong magnetic 
field and potentials that are smooth on the scale of the magnetic
length. In this limit the plateau transition can be described 
by ``quantum percolation'' of semiclassical electron orbits in the 
Chalker-Coddington network model \cite{cc,lwk}.
The latter consists of a square lattice of potential saddle points 
at which quantum tunneling between the edges of quantum Hall droplets 
take place (Fig.~1a). Away from these vertices, the edge electrons 
propagate along the directed links with a fixed chirality set by the 
direction of the magnetic field. To account for the random areas of 
the Hall droplets, the edge electrons accumulate random Bohm-Aharonov 
phases while traversing the links of the network. At each 
saddle point, as shown in Fig.~1b, there are two incoming and two 
out-going edges states. The associated probability amplitudes are 
denoted by $Z_{1,\dots,4}$. Due to current conservation, $\vert 
Z_1\vert^2+\vert Z_2\vert^2 = \vert Z_3\vert^2+\vert Z_4\vert^2$. 
With a choice of gauge, the quantum tunneling
event at each node is then completely specified by a $2\times2$ 
transfer matrix 
\eq
\pmatrix{Z_1\cr Z_3\cr}=
\pmatrix{e^{i\phi_1} & 0\cr
0 & e^{i\phi_2}\cr}
\pmatrix{\cosh\theta & \sinh\theta\cr
\sinh\theta & \cosh\theta\cr}
\pmatrix{e^{i\phi_3} & 0\cr
0 & e^{i\phi_4}\cr}
\pmatrix{Z_4\cr Z_2\cr},
\label{matrix}
\ee
where $\phi_i\in [0,2\pi)$ are random phases, and $\theta$
is a real number. Using Eq.~(\ref{matrix}) as the building
block, we construct the $W\times W$ transfer matrix $T_i$ which 
propagates eigen wavefunctions on a cylinder of circumference $W$ 
in the $x$-direction. The total transfer matrix $T$ for the entire 
$W\times L$ system is given by the matrix product $T=\prod_{i=1}^L
T_i$. For the details of transfer matrix calculations readers are
referred to the original work of Chalker and Coddington \cite{cc}. 
Here we merely emphasize a few important points.

In general the tunneling parameter $\theta$ in
Eq.(\ref{matrix}) should be a random variable depending on the 
local potential at the saddle point. 
However, in Ref.\cite{lwk} it was shown that introducing 
randomness in $\theta$ did not change the universality class
({\it i.e.} did not change the localization length exponent $\nu=2.33\pm0.03$)
of the plateau transition. It turns out that the network model is
situated at the critical point as long as $\theta$ is the same for all 
nodes \cite{lw1}.
Invariance under $90$-degree rotation selects a particular 
one ($\theta=\theta_c=\ln(1+\sqrt{2})$) out of the family 
of critical models \cite{cc,lw1}. 
We shall present results for both isotropic systems ($\theta=\theta_c$)
for which the two-terminal conductance along the $x$-direction ($G_{xx}$) 
and the $y$-direction ($G_{yy}$) are equal, and anisotropic systems 
(i.e. $\theta\ne\theta_c$) for which $G_{xx}\ne G_{yy}$. 
In the latter case we study the geometric mean $\sqrt{G_{xx}G_{yy}}$.

We next define the two-terminal conductance. As shown in Fig.~1a, 
two semi-infinite conducting leads are connected to the $W\times L$ 
disordered network. It has been shown that linear response theory applied 
to the combined lead-sample-lead system gives the following 
multi-channel two-terminal conductance \cite{fisherlee,barastone},
\eq
G={e^2\over h}{\rm Tr} t^\dagger t={e^2\over h}\sum_{i=1}^W{1\over {\rm
    cosh}^2(\gamma_i L)}.
\label{g}
\ee
In Eq.~(\ref{g}), $t$ is the transmission matrix through the 
disordered region and $\gamma_i$ is the $i$-th of the $W$ Lyapunov 
exponents of the hermitian transfer matrix product $T^\dagger T$ 
\cite{pichard}. We note that the validity of Eq.~(\ref{g})
in the absence of time reversal symmetry has been
shown in detail by Baranger and Stone for finite
magnetic fields \cite{barastone}.

We now present the results. We have studied $W\times L$ systems with 
periodic boundary conditions in the transverse direction for 
$L=8,16,32,48,64$, and $96$, and for aspect ratio $W/L=1/4,1/3,1/2,1,2,4$.
The disorder ensemble consists of at least 3000 samples for
each $(W,L)$. 
The distribution function $P(G)$ is very broad, and the most probable
value of the critical conductance $G_{c,{\rm typical}}$ is, although close
to $0.5{e^2\over h}$, not sharply defined.
For isotropic systems (i.e. $\theta=\theta_c$) the 
averaged critical conductance $<G(L)>_c$ for $L\times L$ samples are
shown in Fig.~2. 
To extract the asymptotic value $<G>_c$ and the exponent $y_{\rm irr}$
at the critical point, we expand
the scaling function in Eq.~(\ref{sigma}) according to
${\cal F}(0,x,1)\approx{\cal F}(0,0,1)+{\cal F}'(0,0,1)x$ for small 
$x$. Thus for large system size $L$ we should have
\eq
<G(L)>_c=<G>_c+{\cal F}'(0,0,1)\Delta g_{\rm irr}L^{-y_{\rm irr}}
\label{avgc},
\ee
where $<G>_c={\cal F}(0,0,1){e^2/h}$. From the data shown in Fig.~2, we 
obtain ${\cal F}(0,0,1)=0.58\pm0.03$ and $y_{\rm irr}=0.55\pm0.15$.
We have also studied the aspect-ratio ($W/L$) dependence of $<G>_c$. 
In the context of Eq.~(\ref{sigma}) we find
\eq
{\cal F}(0,0,W/L)= c_1 e^{- c_2 L/W}(W/L),
\ee
with $c_1=0.72\pm0.03$ and $c_2=0.22\pm0.02$.
The exponential factor is a precursor to the conductance behavior 
in a quasi-one dimensional system with $L\gg W$.

In order to verify the universality of $<G>_c$
and to check that the $-y_{\rm irr}$ is indeed the renormalization 
group dimension of the least-irrelevant operator, we have studied network 
models with various forms of node parameter disorder corresponding to 
random distributions of $\theta$-values with the same median
$\theta_c$. The results are consistent within the error bars with 
those obtained above. This further supports the conclusion that 
$\theta$-disorder is an irrelevant perturbation at the critical point. 
In addition, we have also considered anisotropic critical models for 
which $\theta\ne\theta_c$.
In this case, using the procedures described above,
we have calculated the conductances $<G_{xx}>$ and 
$<G_{yy}>$ (periodic boundary conditions are always 
imposed in the transverse directions).

The results are summarized in Table~I. 
Although the critical values $<G_{xx}>_c$ and $<G_{yy}>_c$ depend on 
the amount of anisotropy, their geometric mean $\sqrt{G_{xx}G_{yy}}$ does not. 
The latter stays close to the value of $<G>_c$ 
obtained for the isotropic system. Thus we conclude that there exists 
a {\it universal} critical two-terminal conductance 
$<G>_c=(0.58\pm0.03)e^2/h$.

The fact that our value for $<G>_c$ deviates from $\sxx^c={1\over2}
{{e^2}\over h}$ asserted by KLZ is, as far as we can see, real. 
However, as was pointed out at the beginning of this paper, 
it is not clear how the two-terminal conductance relates to 
the conductivity tensor derived from the Kubo formula in 
systems without contacts.

We now turn to the central moments of the critical conductance
described by the scaling forms given in Eq.~(\ref{sigma2}).
In the inset of Fig.~2 we plot ${\cal F}_2=<(\dg)^2>_c/(e^2/h)^2$
for the isotropic network (i.e. $\theta=\theta_c$) as a
function of the system size for $L\times L$ samples. 
Following a similar finite size scaling analysis as given in
Eq.~(\ref{avgc}), we extract ${\cal F}_2(0,0,1)=0.081\pm0.005$ and the
same $y_{irr}=0.55\pm0.15$ to be the renormalization group dimension 
of the leading irrelevant operator.
The dependence of $<(\dg)^2>_c$ on the degree of anisotropy ({\it i.e.}
$\theta\ne\theta_c$) is shown in Table I. 
Thus we conclude $<(\dg)^2>_c=(0.081\pm0.005)
({e^2\over h})^2$. Repeating the procedure for the 4-th and the 6-th moments
give the results $<(\dg)^4>_c=(0.013\pm0.003)({e^2\over h})^4$,
$<(\dg)^6>_c=(0.0026\pm0.005)({e^2\over h})^6$, and with less accuracy
$<(\dg)^8>_c=(8\pm2)\times10^{-4}({e^2\over h})^8$.
In Fig.~3, we present the data obtained for even-integer as well
as odd-integer and non-integer central moments.
The $n$-th order moments interpolate between
$<(\dg)^n>_c=a v^n e^{un^2}$ at small $n$ and
$<(\delta G)^n>_c=bn^{-\beta}$ at large $n$, where
$(a,v,u,b,\beta)\approx(0.80,0.28,0.054,4.65,4.18)$ are
all universal constants. The small-$n$ behavior shows that {\it the
critical conductance obeys a log-normal distribution}. 
On the other hand, the large-$n$ behavior, indicative of a 
broad distribution, is purely empirical.
The latter should be
contrasted to the behavior in mesoscopic disordered metals where 
the the $2+\epsilon$-expansion in the diffusive regime predicts
nonuniversal moments for large $n$ \cite{lerner}.

We now compare our results with experimental findings. 
This comparison must be made under the {\it disclaimer} that so far 
the understanding of the effects of Coulomb interaction on the critical 
properties of the integer plateau transition is still incomplete \cite{lw2}.
First, we {\it assume} that the two-terminal conductance $<G>_c$ is 
the four-probe $\sxx^c$ measured experimentally. Second, from general 
particle-hole symmetry argument \cite{klz} one expects
$\sxy^c=0.5{e^2\over h}$. With our result $<G>_c\approx 0.58 {e^2\over h}$,
this implies $\rho_{xx}^c=0.99 {h\over e^2}$. This value agrees
with the experimental finding of Shahar {\it et. al.} \cite{shahar}.
More recently, Cobden and Kogan have measured the 
two-terminal conductance near the integer transitions.
\cite{cobden}. Since the latter was carried out on mesoscopic
samples outside the asymptotic scaling regime, direct comparison
with our results would be difficult to justify. Nevertheless,
their data provide an estimated value $<(\delta G)^2>\approx 
0.05(e^2/h)^2$, which is in reasonable agreement with our findings.

{\noindent Acknowledgment:} We thank A.~D. Stone and X.-G. Wen
for useful discussions, and D.~H. Cobden for sending us his 
experimental data before publication. 
ZW acknowledges the support of Research Corporation.

\begin{figure}
Figure 1. (a) Schematics of the network model. 
The shaded areas correspond to the semi-infinite metallic leads.
(b) The quantum tunneling at a single node of the network.
\label{fig1}
\end{figure}
\epsfsize = 17cm
\rotate[r]{\epsffile{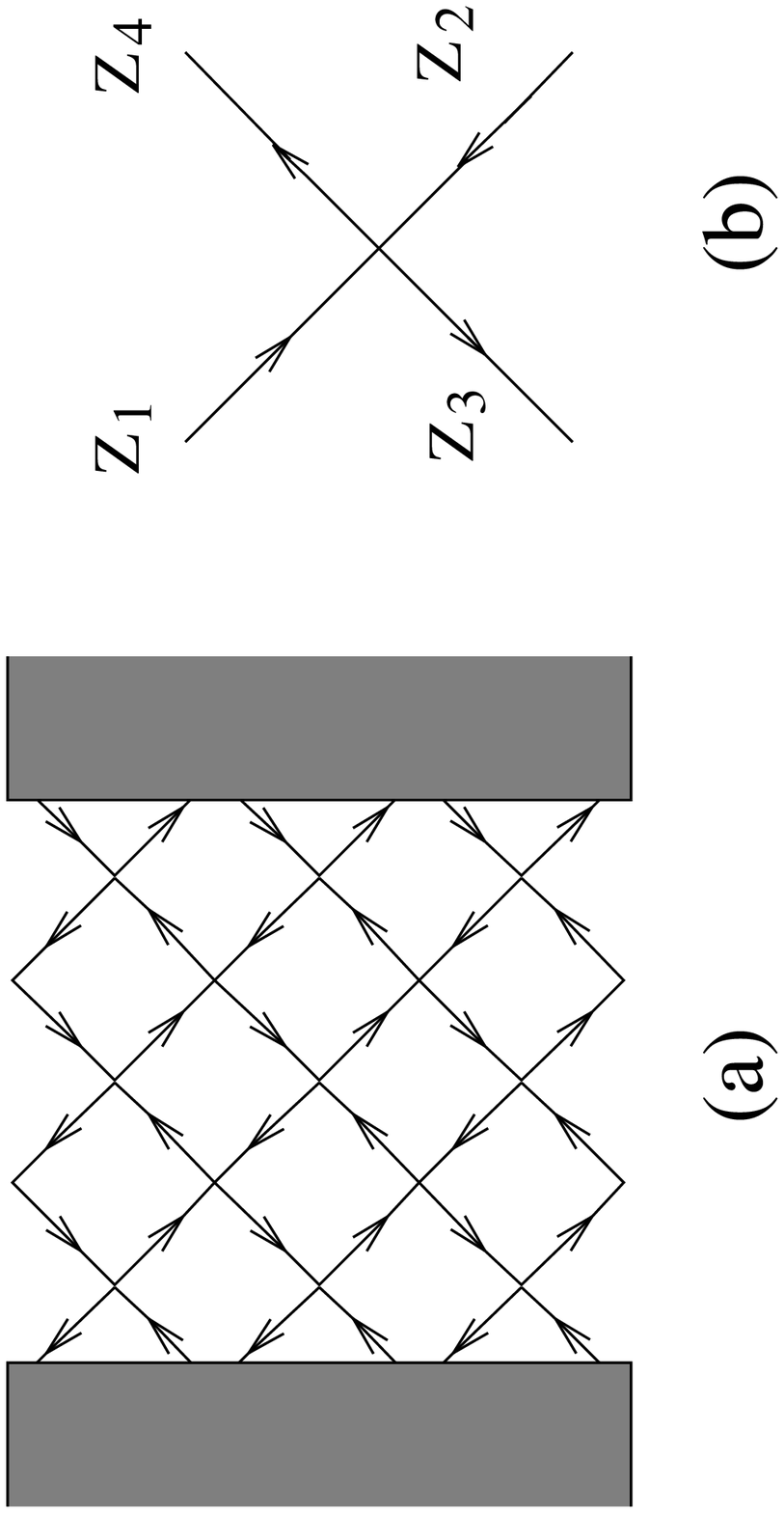}}

\vskip 20pt
\begin{figure}
Figure 2. The finite-size dependence of the critical conductance at the
isotropic critical point. The solid line is the fit to the scaling 
form in Eq.~(\ref{avgc}) in the text with
$<G>_c=(0.58\pm0.03)e^2/h$ and $y_{\rm irr}=0.55\pm0.15$. 
Inset shows the behavior of the conductance
fluctuations, which gives
$<(\dg)^2>_c=(0.081\pm0.005)(e^2/h)^2$ and the same renormalization
group dimension $y_{\rm irr}$.
\label{fig2}
\end{figure}
\epsfsize = 17cm
\rotate[r]{\epsffile{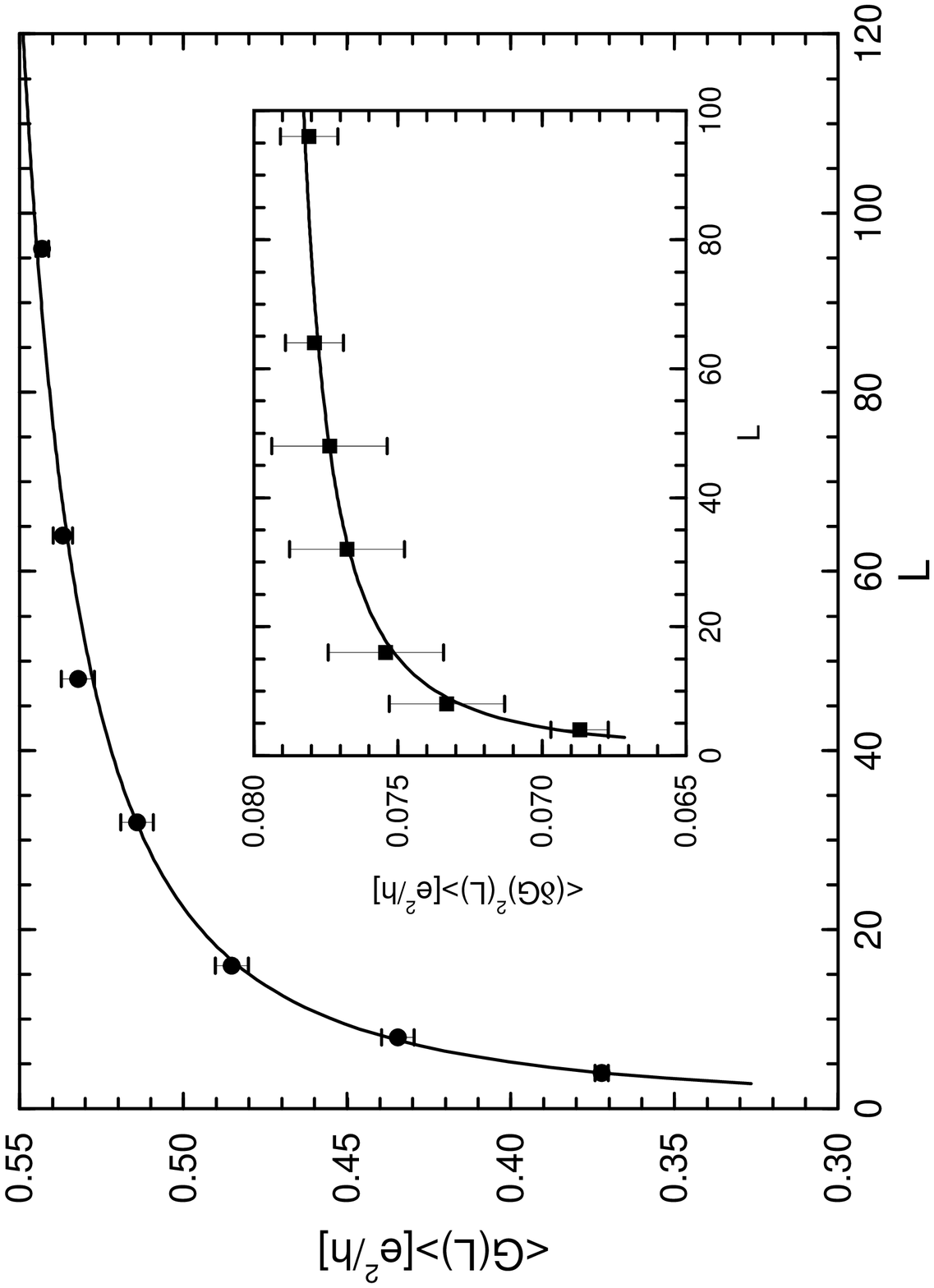}}

\newpage
\begin{figure}
Figure 3. The behavior of the $n$-th order central moments of
the conductance distribution which interpolates 
between the exponential and the power-law
behaviors (dashed-lines) in the small and large $n$ limits 
as described in the text.
\label{fig3}
\end{figure}
\epsfsize = 17cm
\rotate[r]{\epsffile{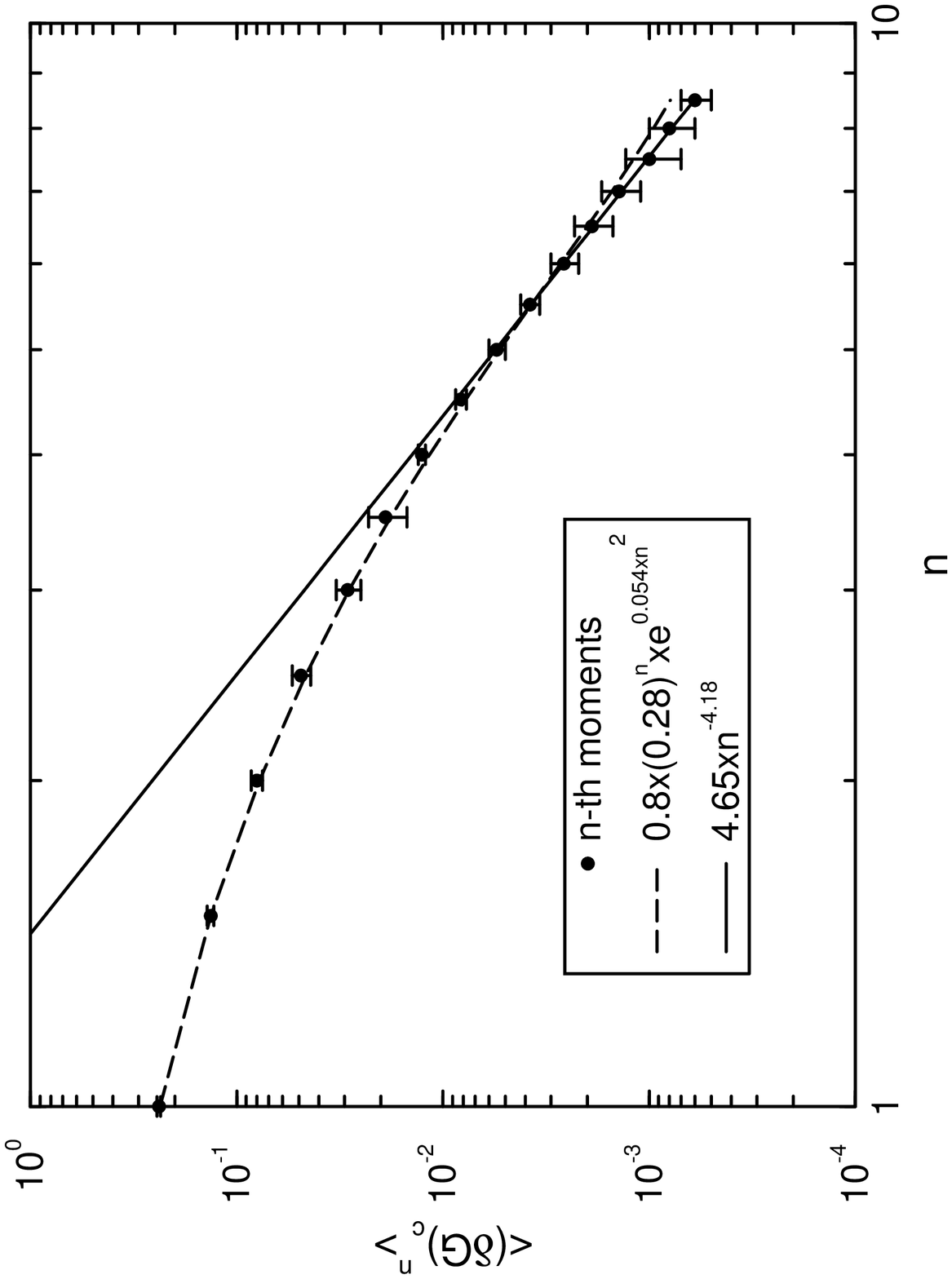}}

\begin{table}
\caption{Ensemble-averaged conductances and conductance fluctuations 
  at the critical points of the isotropic ($\theta=\theta_c$) and the 
  anisotropic ($\theta\ne\theta_c$) models. The units are $e^2/h$ for 
  $G$ and $(e^2/h)^2$ for $(\dg)^2$. Error estimates in the last digits 
  are given in parenthesis.}
\begin{tabular}{c||ccc|ccc}
      $\theta$ & $G_{xx} $ & $G_{yy}$ &
      $\sqrt{G_{xx}G_{yy}}$ \  & $(\dg_{xx})^2$ & $(\dg_{yy})^2$ &
      $\sqrt{(\dg_{xx})^2(\dg_{yy})^2}$ \\
\hline
$\theta_c$ & $0.58(3)$& $0.58(3)$ & $0.58(3)$ \  & $0.081(5)$ & $0.081(5)$
& $0.081(5)$ \\
\hline
$0.84$ & $0.64(3)$ & $0.53(3)$ & $0.58(3)$ \ & $0.081(5)$ & $0.084(5)$
& $0.082(5)$ \\
$0.80$ & $0.75(3)$ & $0.47(3)$ & $0.59(3)$ \ & $0.080(5)$ & $0.086(3)$
& $0.082(4)$ \\
$0.75$ & $0.88(3)$ & $0.43(3)$ & $0.62(4)$  \ & $0.075(5)$ & $0.083(5)$
& $0.079(5)$ \\
$0.70$ & $1.04(3)$ & $0.30(3)$ & $0.56(4)$ \  & $0.078(6)$ & $0.079(5)$
& $0.079(6)$
\end{tabular}
\label{tab1}
\end{table}


\begin{references}
\bibitem{footnote} Averaging over a disordered ensemble in this paper
mimics the self-averaging when the phase coherence length is 
smaller than the sample dimension.
\end{references}
\end{document}